\documentclass[11pt, a4paper]{article}

\usepackage{proceedings}   

\confname{The 21st International Conference on B-Physics at Frontier Machines, BEAUTY 2026}
\confdate{1--5 June 2026}
\conflocation{Maastricht, Netherlands}
\confLogo[width=8cm]{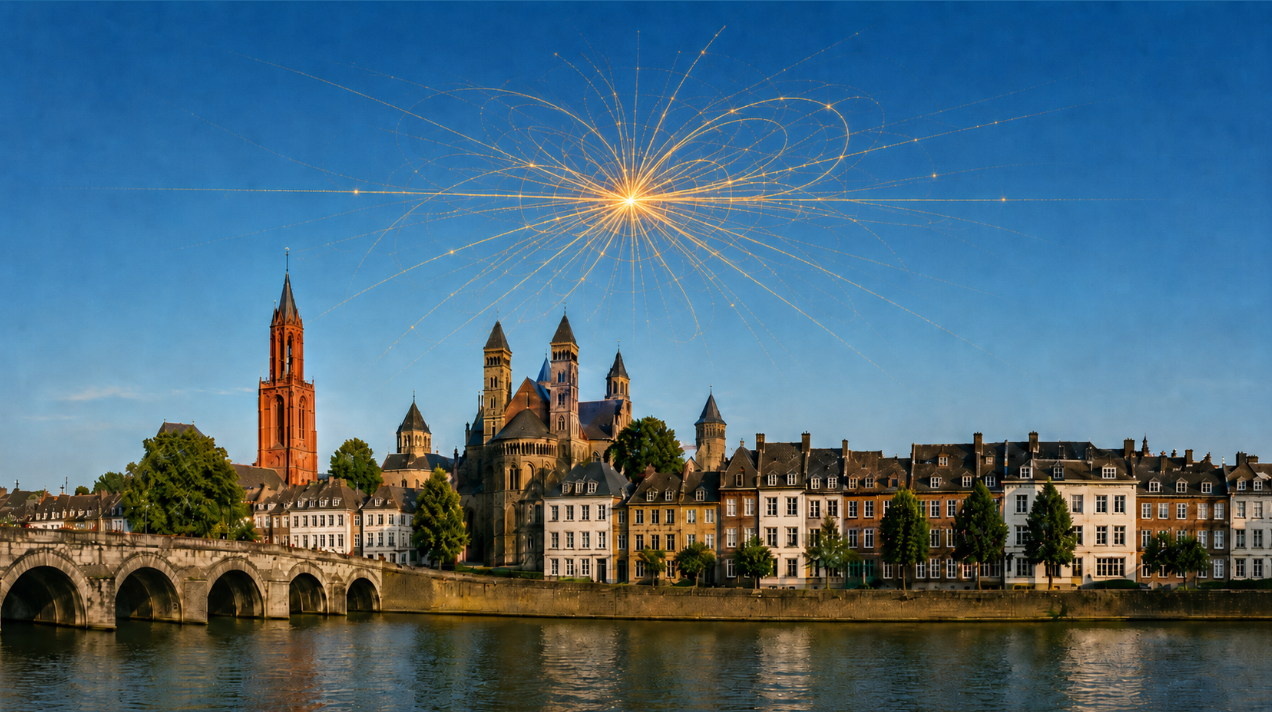} 

\title{Born-Oppenheimer Effective Field Theory as a unified framework for all the XYZ exotics}

\author{%
  Tommaso Scirpa\textsuperscript{1,*}
}

\affiliation[1]{Technical University of Munich,\\
TUM School of Natural Sciences, Physics
Department,\\   
James-Franck-Strasse~1, 85748 Garching, Germany.}

\email{tommaso.scirpa@tum.de}

\usepackage{physics}
\usepackage{slashed}
\usepackage{xspace}
\usepackage{siunitx}
\usepackage[numbers,sort&compress]{natbib}

\begin{document}

\maketitle
\printaffiliations
\vspace{0.3cm}

\begin{abstract}
We discuss some recent applications of the Born-Oppenheimer effective field theory (BOEFT) formalism, directly derived from QCD and grounded in scale separation and symmetries, to hadronic systems containing two heavy quarks.
This formalism provides a unified description of both ordinary and exotic quarkonium states (i.e., hybrids, tetraquarks, pentaquarks, ...) dynamics via coupled Schrödinger equations, whose potentials are computable on the lattice.
The aim is to unveil the elusive pattern of the XYZ exotic hadrons.
Here, we apply the BOEFT formalism to the $\chi_{c1}(3872)$, $T_{cc}(3875)^+$, and their bottom counterparts. Moreover, we assess the threshold effects on the quarkonium spectrum and discuss the inclusive production cross section for $\chi_{c1}(3872)$.
\end{abstract}


\section{Introduction}
\label{sec:intro}

Since the early days of QCD, the possibility of color-neutral combinations of quarks and gluons different from the ordinary mesons or baryons was conceived~\cite{Gell-Mann:1964ewy,Zweig:1964ruk,Jaffe:1975fd}.
The first experimental evidence, however, came much later with the discovery of the $\chi_{c1}(3872)$ by Belle in 2003~\cite{Belle:2003nnu}, followed by tens of additional states observed at the B-factories, $\tau$-charm factories, and hadron colliders (see~\cite{Esposito:2016noz,Guo:2017jvc,Ali:2017jda,Brambilla:2019esw} for reviews, and \url{https://qwg.ph.nat.tum.de/exoticshub/} for an up-to-date list of discoveries).
Most of these exotic states contain two heavy quarks, appear in the charmonium and bottomonium sectors, and have collectively been dubbed \textit{XYZ states}.
Their exotic nature manifests in spectroscopic properties (mass, decay patterns, quantum numbers, \ldots) inconsistent with conventional quarkonium, pointing instead to more complex multiquark structures such as hybrids, tetraquarks, and pentaquarks.
Their debated internal structure has led to the formulation of phenomenological models relying on specific assumptions about the dominant degrees of freedom, such as hadronic molecules, compact tetraquarks, hadroquarkonium, or other multiquark states.
While successful in describing individual candidates, these models fail to capture the broader picture of the XYZ states and lack a direct connection to QCD.



\section{The BOEFT framework}
\label{sec:boeft}

The BOEFT describes exotic hadrons made of two heavy quarks (the heavy degrees of freedom (HDF)), $Q\bar{Q}$ or $QQ$ with $Q = c,\,b$, plus light degrees of freedom (LDF) -- light quarks or gluons -- in a color-singlet state, with characteristic energy scales $m_Q$ (heavy-quark mass), $m_Q v$ (momentum), $\Lambda_{QCD}$ (LDF energy), and $m_Q v^2$ (binding energy), hierarchically ordered as $m_Q \gg m_Q v \sim 1/r \gtrsim \Lambda_{QCD} \gg m_Q v^2$ since the heavy quarks move nonrelativistically ($v \ll 1$), with $r$ the heavy-quarks separation.
This hierarchy yields a tower of effective field theories (EFTs) down to the binding-energy scale. The last step, integrating out scales up to $m_Q v^2$, is the BOEFT~\cite{Berwein:2015vca,Oncala:2017hop,Brambilla:2017uyf,Soto:2020xpm,Berwein:2024ztx}, generalizing pNRQCD~\cite{Brambilla:1999xf,Brambilla:2004jw} -- originally developed for quarkonium -- to any XYZ exotic states. Integrating out $\Lambda_{QCD}$ exploits the time-scale separation $\tau_{\rm HDF} \sim 1/m_Q v^2 \gg \tau_{\rm LDF} \sim 1/\Lambda_{QCD}$, so that the LDF adjust (almost) instantaneously to the heavy-quark motion. This is the Born--Oppenheimer approximation from molecular physics~\cite{Born:1927rpw}, incorporated at leading order in the BOEFT.

In the static limit ($m_Q \rightarrow \infty$), $Q\bar{Q}$ or $QQ$ pairs act as color sources and different exotic systems, distinguished by their LDF, generate distinct static potentials encoding the nonperturbative dynamics. The QCD symmetry group breaks down to the cylindrical group $D_{\infty h}$, whose irreducible representations $\Lambda_\eta^\sigma$ -- the \textit{BO quantum numbers} -- classify these potentials. They are defined as follows: $\Lambda = |\bm{K} \cdot \hat{\bm{r}}|$ ($\bm{K}$ the LDF angular momentum, $\hat{\bm{r}}$ the heavy-quark axis unit vector), denoted $\Sigma, \Pi, \Delta, \ldots$ for $\Lambda = 0, 1, 2, \ldots$; $\eta = +1\,(g), -1\,(u)$, the $PC$ ($P$) eigenvalue for $Q\bar{Q}$ ($QQ$); and $\sigma = \pm 1$, the LDF reflection eigenvalue, given only when $\Lambda = 0$. Potentials sharing the same quantum numbers are further distinguished by primes, $\Lambda_\eta^\sigma, \Lambda_\eta^{\sigma\prime}, \Lambda_\eta^{\sigma\prime\prime}, \ldots$ (Tables~\ref{tab:multiplets},~\ref{tab:QQqq}). As $r \rightarrow 0$, the cylindrical group enlarges to the spherical one $O(3)$ ($\times C$) for $QQ$ ($Q\bar{Q}$), with representations labeled by $k^{P(C)}$ ($k(k+1)$ are the $\bm{K}^2$ eigenvalues), so all the potentials $\Lambda_\eta^\sigma$ with the same $k^{P(C)}$ become degenerate at small $r$ (Figure~\ref{fig:potentials}). The LDF states are thus classified by $\kappa = \{k^{P(C)}, f\}$, with $f$ the flavor quantum numbers (e.g., isospin, baryon number).

The BOEFT Lagrangian, obtained by matching to NRQCD and integrating out modes with $m_Q v \gtrsim \Lambda_{QCD}$, takes the following form at leading order in the $v$ expansion~\cite{Berwein:2024ztx}

\begin{align}\label{eq:BOEFTlag}
L_{\rm BOEFT} = \int d^3\boldsymbol{R} \int d^3\boldsymbol{r} \sum_{\kappa\lambda\lambda'} \mathrm{Tr}&\left\{ \Psi^\dagger_{\kappa\lambda}(\boldsymbol{r},\boldsymbol{R},t) \left[ i\partial_t\,\delta_{\lambda\lambda'} - V_{\kappa,\lambda\lambda'}(r) \right. \right. \\
&+ \left. \left. \sum_\alpha P^{\alpha\dagger}_{\kappa\lambda}(\theta,\varphi)\, \frac{\boldsymbol{\nabla}^2_r}{m_Q}\, P^\alpha_{\kappa\lambda'}(\theta,\varphi) \right] \Psi_{\kappa\lambda'}(\boldsymbol{r},\boldsymbol{R},t) \right\},\nonumber
\end{align}
where the trace runs over color, spin, and isospin indices, and $\mathbf{R}$ is the heavy-quarks center-of-mass coordinate. $\Psi_{\kappa\lambda}$ denotes the exotic field with LDF quantum numbers $\kappa,\lambda$, and $P^\alpha_{\kappa\lambda'}$ the corresponding projectors. The potentials $V_{\kappa,\lambda\lambda'}$ can be expressed as generalized Wilson loops, computable on the lattice order by order in the $1/m_Q$ expansion,
\begin{align}
V_{\kappa\lambda\lambda'}(r) = V^{(0)}_{\kappa,|\lambda|}(r)\delta_{\lambda\lambda'} + \frac{V^{(1)}_{\kappa\lambda\lambda'}(r)}{m_Q} + \cdots
  \label{eq:BOEFTpot},
\end{align}
and encode the nonperturbative dynamics of the LDF.
$V^{(0)}_{\kappa,|\lambda|}(r)$ are the static potentials (denoted by $V_{\Lambda_\eta^\sigma}(r)$ hereafter), while $V^{(1)}_{\kappa\lambda\lambda'}$ encode the leading spin-dependent contributions.
The coupled Schrödinger equation relative to Eq.~\eqref{eq:BOEFTlag} is 
\begin{align}
\sum_{\lambda} \left[ -\frac{1}{m_Q r^2} \partial_r r^2 \partial_r + \frac{1}{m_Q r^2} M_{\lambda'\lambda} + V_{\kappa\lambda'\lambda} \right] \psi^{(N)}_{\kappa\lambda}(r) = \mathcal{E}_N \, \psi^{(N)}_{\kappa\lambda'}(r),
\label{eq:schr}
\end{align}
where the states' quantum numbers are collectively denoted $N \equiv \{n, j, m_j, l, s\}$: $n$ is the principal quantum number, the rest are eigenvalues of $\mathbf{J}^2$, $J_3$, $\mathbf{L}^2$, $\mathbf{S}^2$, with $\mathbf{S} = \mathbf{S}_Q + \mathbf{S}_{\bar{Q}(Q)}$, $\mathbf{L} = \mathbf{L}_{Q} + \mathbf{K}$, $\mathbf{J} = \mathbf{L} + \mathbf{S}$, with $\mathbf{S}_Q$/$\mathbf{S}_{\bar{Q}}$, $\mathbf{L}_{Q}$ being the heavy (anti)quark spin, and the heavy-quarks angular momentum. The angular momentum matrix $M_{\lambda'\lambda}$ has dimension $\min(2k+1,2l+1)$, and can be decomposed into blocks of defined parity. 

Besides short-distance mixing from spherical-symmetry restoration, BO potentials may mix in another way: when two potentials sharing BO quantum numbers come close over some $r$ range, their adiabatic potentials undergo an avoided level crossing (see Figure~\ref{fig:potentials}). This mechanism drives the quarkonium--lowest-tetraquark-potential mixing via string breaking, with strong implications for both $\chi_{c1}(3872)$ and the quarkonium spectrum (see Sections~\ref{sec:XandTcc},~\ref{sec:quarkonium}).

\section{$\chi_{c1}(3872)$, $T_{cc}(3875)^+$ and their bottom partners in BOEFT}
\label{sec:XandTcc}

We introduce the coupled Schrödinger equations relevant for the $\chi_{c1}(3872)$ and $T_{cc}(3875)^+$, constraining the potentials to the available lattice data. The only free parameters of the model, the adjoint and triplet meson masses, are tuned to generate shallow bound states below threshold, corresponding to $\chi_{c1}(3872)$ and $T_{cc}(3875)^+$~\cite{Belle:2003nnu,LHCb:2021vvq, LHCb:2021auc,ParticleDataGroup:2024cfk}.
In the BOEFT description, no specific internal structure (e.g., compact tetraquark or hadronic molecule) is assumed for these states; rather, they emerge from the interplay between the short- and long-range behavior of the tetraquark potentials.

\subsection{$\chi_{c1}(3872)$ and $X_b$}
\label{subsec:X}

The $\chi_{c1}(3872)$ is an isoscalar state and a possible $c\bar{c}q\bar{q}$ tetraquark candidate ($q = u, d$), with $J^{PC} = 1^{++}$, lying about $100\,\mathrm{keV}$ from the $D^{0*}\bar{D}^0$ threshold~\cite{ParticleDataGroup:2024cfk}. Here, we provide a leading-order BOEFT description of the state, accounting for both the quarkonium and the lowest-lying tetraquark potentials, which mix due to string breaking.

Symmetry arguments constrain these potentials' asymptotic behavior. As $r \rightarrow 0$, the $Q\bar{Q}$ pair is a static color source, in either a singlet $(Q\bar{Q})_1$ or an octet $(Q\bar{Q})_8$ representation. For $(Q\bar{Q})_1$, the LDF carry the vacuum quantum numbers $k^{PC} = 0^{++}$, corresponding to $\Sigma_g^+$ and giving the quarkonium potential, which exhibits an attractive Coulombic behavior at short distances.
For $(Q\bar{Q})_8$, the LDF are in this case \textit{adjoint mesons} $(q\bar{q})_8$, giving $[(Q\bar{Q})_8 + (q\bar{q})_8]$ tetraquarks. The lowest tetraquark potentials, whose \textit{adjoint meson} has $k^{PC} = 1^{--}$, and  $\Sigma_g^{+\prime}$ and $\Pi_g$ BO quantum numbers, exhibit a repulsive Coulombic behavior at short distances, with an offset given by the adjoint meson mass.\footnote{There is also another tetraquark potential,  whose adjoint meson has quantum numbers $k^{PC} = 0^{-+}$,   corresponding to $\Sigma_u^-$, and a mass comparable to that of $k^{PC} = 1^{--}$~\cite{Foster:1998wu}. This potential is not discussed further here, as it does not enter the coupled Schr\"odinger equation relevant to the $\chi_{c1}(3872)$ state.} These two potentials mix at short distances due to spherical-symmetry restoration. At large distances, instead, they both approach an S-wave heavy-light meson-antimeson configuration, $[(Q\bar{q})_1 + (\bar{Q}q)_1]$, whose LDF quantum numbers coincide with those of the lowest adjoint mesons. Due to the \textit{$\Lambda_\eta^\sigma$ exact conservation in the static limit at all distances}, the repulsive color-octet behavior at short distances must connect smoothly with the flat behavior associated with the meson-antimeson configuration~\cite{Berwein:2024ztx,Braaten:2024tbm,Brambilla:2024imu}. This transition is governed by pion-exchange dynamics.
This shape lets tetraquark potentials support only finitely many bound-state or resonant solutions, naturally located near threshold. Their existence is set by the adjoint meson mass value, which determines whether the potentials approach the threshold from below or above, thereby allowing for one or more such solutions, or none.

\begin{figure}[h]
  \centering
  \includegraphics[width=0.5\linewidth]{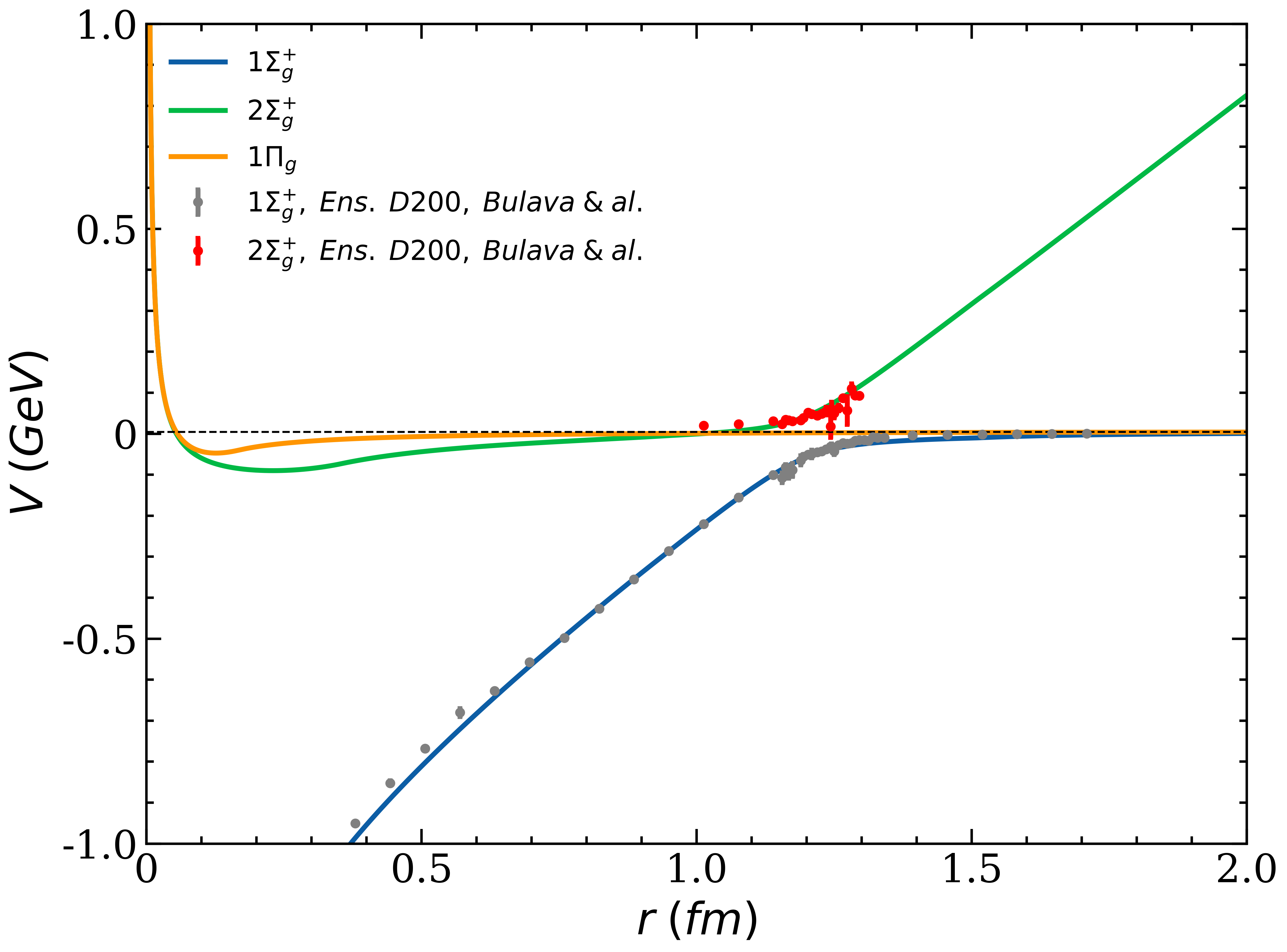}
  \caption{The adiabatic QCD potentials of the system~\cite{Brambilla:2026jqz}. Note the short-distance degeneracy between the $2\Sigma_g^+$ and $1\Pi_g$ static potentials, due to spherical symmetry restoration, and the narrow avoided
   crossing between the $1\Sigma_g^+$ and $2\Sigma_g^+$ potentials, due to the quarkonium-tetraquark mixing via string breaking.  The lattice QCD data points are relative to the D200 ensemble of Ref.~\cite{Bulava:2024jpj}.}
  \label{fig:potentials}
\end{figure}

The lowest $J^{PC}$ multiplets associated with the $\Sigma_g^+$ quarkonium and $\Sigma_g^{+\prime}$, $\Pi_g$ tetraquark potentials are given in Table~\ref{tab:multiplets}. We identify the $\chi_{c1}(3872)$ as the $1^{++}$ state of the lowest multiplet of $k^{PC} = 1^{--}$. 
The BOEFT coupled Schrödinger equation accounting for the mixing terms mentioned above is~\cite{Brambilla:2024imu,Brambilla:2026jqz}:

\begin{table}[h!]
\centering
\resizebox{0.70\columnwidth}{!}{%
	\begin{tabular}{||c|c|c||c|c|c||}
		\hline\hline
		\multirow{2}{*}{\hspace{2pt}$\begin{array}{c} Q\bar{Q}\\\text{color state}\end{array}$\hspace{2pt}} & \multirow{2}{*}{\hspace{2pt}$\begin{array}{c}\text{$q\bar{q}$ spin}\\k^{PC}\end{array}$\hspace{2pt}} & \multirow{2}{*}{\hspace{2pt} $\begin{array}{c} \text{BO quantum \#}\\\Lambda^\sigma_\eta\end{array}$\hspace{2pt}}&\multirow{2}{*}{\hspace{2pt} $l$\hspace{2pt}}& \multirow{2}{*}{\hspace{2pt} $\begin{array}{c}J^{PC}\\\{S=0, S=1\}\end{array}$\hspace{2pt}}& \multirow{2}{*}{\hspace{2pt}Multiplets\hspace{2pt}}\\
		& & & & &  \\
		\hline\hline
		\multirow{6}{*} {\hspace{2pt}$\begin{array}{c} \text{Octet}\\\mathbf{8}\end{array}$\hspace{2pt}}
		&\multirow{3}{*}{$0^{-+}$} & \multirow{3}{*}{$\Sigma_u^-$}  & \hspace{2pt}$0$\hspace{2pt} & \hspace{2pt}$\{0^{++}, 1^{+-}\}$\hspace{2pt}
		&\hspace{2pt}$T_1^0$\hspace{2pt}\\
		\cline{4-6}
		& & & \hspace{2pt}$1$\hspace{2pt} & \hspace{2pt}$\{1^{--}, \left(0, 1, 2\right)^{-+}\}$\hspace{2pt}&\hspace{2pt}$T_2^0$\hspace{2pt}\\
		\cline{2-5}\cline{2-6}
		&\multirow{4}{*}{$1^{--}$} & ${\Sigma_g^{+\prime},\Pi_g}$  & \hspace{2pt}$1$\hspace{2pt} & \hspace{2pt}$\{1^{+-}, (0,1,2)^{++}\}$\hspace{2pt}&\hspace{2pt}$T_1^1$\hspace{2pt} \\
		\cline{3-6}
		& & ${\Sigma_g^{+\prime}}$  & \hspace{2pt}$0$\hspace{2pt} & \hspace{2pt}$\{0^{-+},  1^{--}\}$\hspace{2pt}&\hspace{2pt}$T_2^1$\hspace{2pt} \\
		\cline{3-6}
		& & ${\Pi_g}$  & \hspace{2pt}$1$\hspace{2pt} & \hspace{2pt}$\{1^{-+},  (0,1,2)^{--}\}$\hspace{2pt}&\hspace{2pt}$T_3^1$\hspace{2pt} \\
		\cline{3-6}
		\hline\hline
	\end{tabular}}
	\caption{$J^{PC}$ multiplets for quarkonium $Q \bar{Q}$ and the lowest tetraquarks $Q \bar{Q} q \bar{q}$ with $l\leq1$.
    The multiplets are ordered by increasing energies of the relative states.}
	\label{tab:multiplets}
\end{table}


\begin{equation}
\scalebox{0.85}{$
\begin{aligned}
&\left[
-\frac{1}{m_Qr^2}\,\partial_r r^2\partial_r+\frac{1}{m_Qr^2}
\begin{pmatrix}
l\left(l+1\right) & 0 & 0\\[6pt]
0 & l(l+1)+2 & -2\sqrt{l(l+1)} \\[6pt]
0 & -2\sqrt{l(l+1)} & l(l+1)
\end{pmatrix}
\right.\\
&\hspace{4.0 cm}\left.
+\begin{pmatrix}
V_{\Sigma_{g}^{+}}(r) & V_{\Sigma_{g}^{+}-\Sigma_{g}^{+\prime}}(r) & 0 \\[6pt]
V_{\Sigma_{g}^{+}-\Sigma_{g}^{+\prime}}(r) & V_{\Sigma_{g}^{+\prime}}(r) & 0\\[6pt]
0 & 0 & V_{\Pi_g}(r)
\end{pmatrix}
\right]
\begin{pmatrix}
\psi_{\Sigma_g^+}^{(N)}(r) \\[6pt]
\psi_{\Sigma_g^{+\prime}}^{(N)}(r) \\[6pt]
\psi_{\Pi_g}^{(N)}(r)
\end{pmatrix}
=
\mathcal{E}_{N}
\begin{pmatrix}
\psi_{\Sigma_g^+}^{(N)}(r) \\[6pt]
\psi_{\Sigma_g^{+\prime}}^{(N)}(r) \\[6pt]
\psi_{\Pi_g}^{(N)}(r)
\end{pmatrix}.
\end{aligned}
$}
\label{eq:coupledSchr}
\end{equation}

where the $V_{\Sigma_{g}^{+}}$, $V_{\Sigma_{g}^{+\prime}}$, $V_{\Sigma_{g}^{+}-\Sigma_{g}^{+\prime}}$ are constrained from the available lattice data~\cite{Bulava:2024jpj}.
Figure~\ref{fig:potentials} displays the adiabatic potentials of the system, obtained by diagonalizing the potential matrix of Eq.~\eqref{eq:coupledSchr}.
The only free parameter, the $k^{PC} = 1^{--}$ adjoint meson mass relative to the $V_{\Sigma_{g}^{+\prime}}$, $V_{\Pi_g}$ potentials, has been tuned to generate a single shallow bound state solution, identified with the $\chi_{c1}(3872)$.

\begin{figure}[htbp]
    \centering
    \begin{subfigure}[b]{0.44\textwidth}
        \centering
        \includegraphics[width=\textwidth]{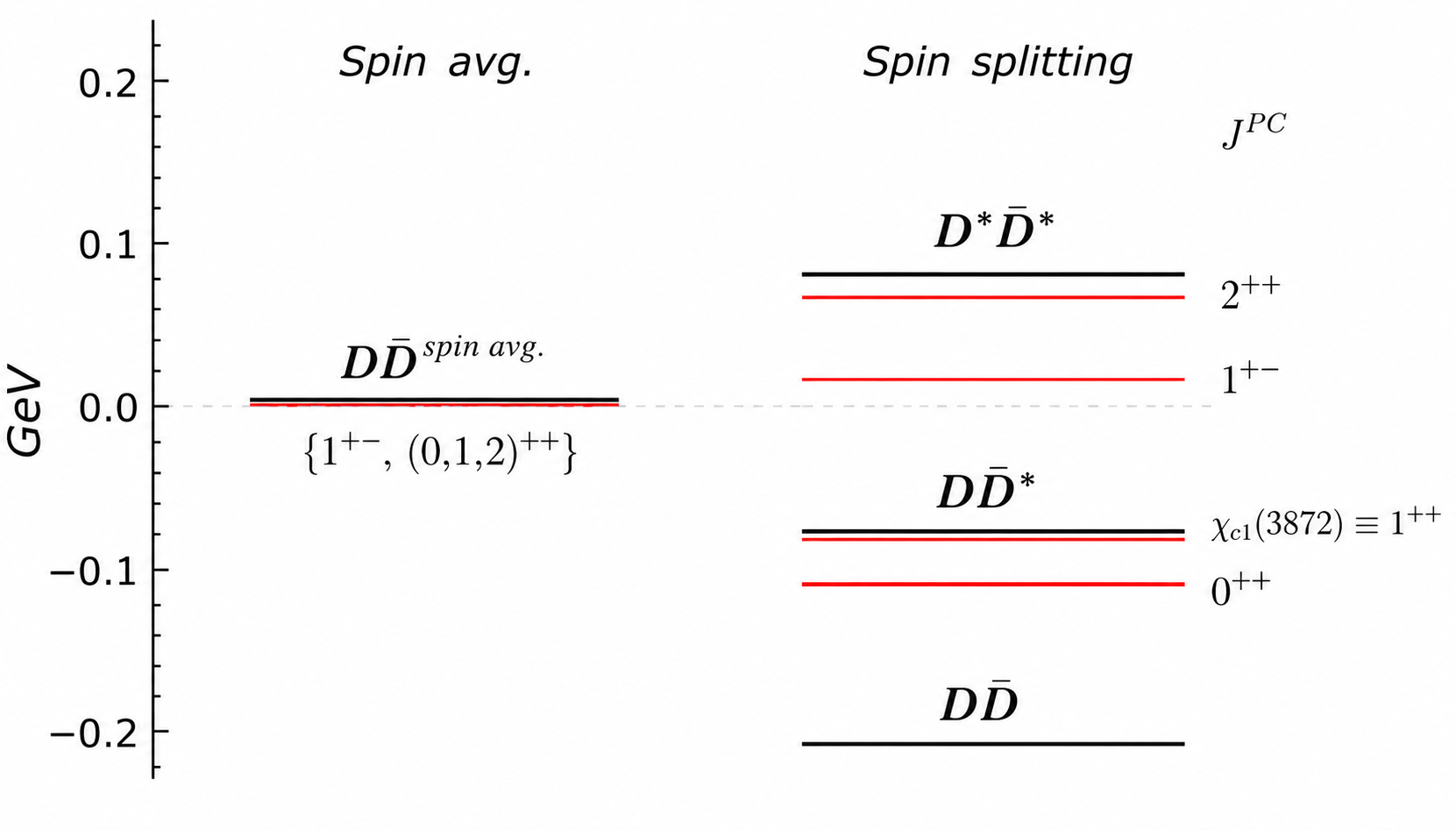}
    \end{subfigure}
    \hfill
    \begin{subfigure}[b]{0.44\textwidth}
        \centering
        \includegraphics[width=\textwidth]{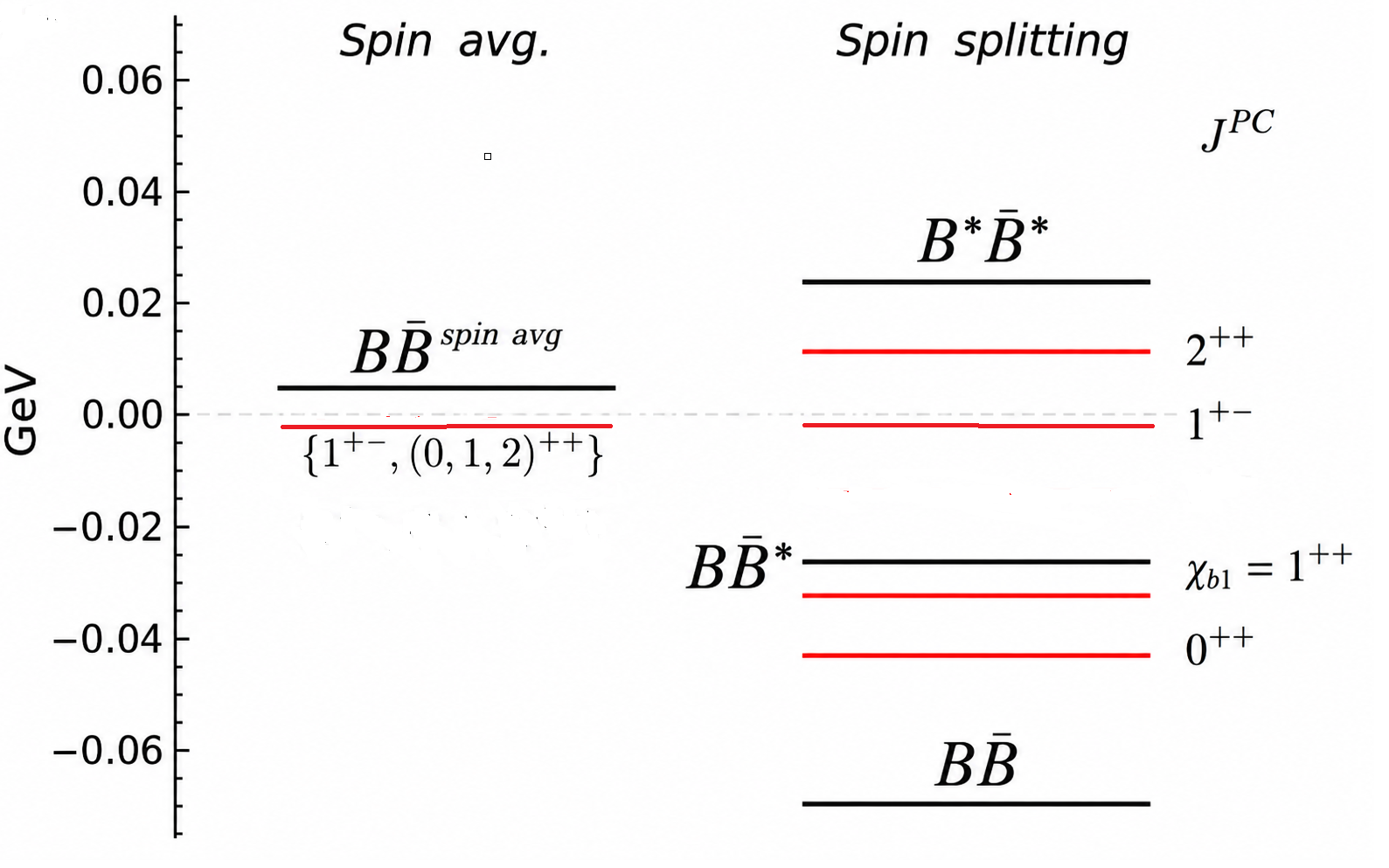}
    \end{subfigure}
    \caption{Mass spectrum for the $\chi_{c1}(3872)$ multiplet (left panel) and its bottom counterpart $X_b$ (right panel)~\cite{Brambilla:2024imu}. 
    The meson-antimeson thresholds are indicated with black lines, while the BO states are indicated with red ones.}
    \label{fig:spin-splitting-multiplets}
\end{figure}

The resulting $\chi_{c1}(3872)$ state has a large scattering length $a \sim 15$~fm and substantial tetraquark component $|\psi_{Q\bar{Q}q\bar{q}}|^2 \sim 92 \%$, along with a residual quarkonium component $|\psi_{Q\bar{Q}}|^2 \sim 8 \%$.
Assuming that the radiative decays $\chi_{c1}(3872) \rightarrow J/\psi \, \gamma$, $\chi_{c1}(3872) \rightarrow \psi(2S) \, \gamma$ proceed through the quarkonium component, and fixing their values from quark models~\cite{Pei:2024hzv}, we get
\begin{equation}\label{eq:ratio}
\mathcal{R}_{\gamma \psi} = \frac{\Gamma_{\chi_{c1}(3872) \rightarrow \psi(2S) \, \gamma}}{\Gamma_{\chi_{c1}(3872) \rightarrow J/\psi \, \gamma}} = 2.99 \pm 2.36,
\end{equation}
consistent with the LHCb value of $1.67 \pm 0.25$~\cite{LHCb:2024tpv}.
Accounting for the $\mathcal{O}(1/m_Q)$ tetraquark spin-dependent corrections, we predict the $\chi_{c1}(3872)$ mass being $3872(14)$~MeV, just below the $D \bar{D}^*$ threshold. The other members of the multiplet have masses of $3957(11)$~MeV for the $1^{+-}$ state, $3846(11)$~MeV for the $0^{++}$ state, and $4004(14)$~MeV for the $2^{++}$ state (see Figure~\ref{fig:spin-splitting-multiplets}).
In our model, the $\chi_{c1}(3872)$ state is different from the   $\chi_{c1}(2P)$ quarkonium state, for which we predict a mass a few MeV above the $D \bar{D}^*$ threshold~\cite{Brambilla:2026zfw}. 

Solving Eq.~\eqref{eq:coupledSchr} for the bottom sector, with the adjoint meson mass tuned on the $\chi_{c1}(3872)$, gives a state $X_b$ a few MeV below the corresponding $B\bar{B}^*$ threshold, consisting of a $98\%$ tetraquark component and a residual $2\%$ quarkonium one. The full multiplet structure is shown in Figure~\ref{fig:spin-splitting-multiplets}. The same approach has also been applied to describe the hidden-charm pentaquark spectrum -- $P_{c\bar{c}}(4312)^+$, $P_{c\bar{c}}(4380)^+$, $P_{c\bar{c}}(4440)^+$, $P_{c\bar{c}}(4457)^+$ -- by tuning the relevant adjoint baryon masses, and to predict the masses of their bottom counterparts~\cite{Brambilla:2025xma}.

\subsection{$T_{cc}(3875)^+$, $T_{bb}$ and $T_{bc}$}
\label{subsec:T}

The $T_{cc}(3875)^+$ is the first doubly heavy tetraquark observed at LHCb~\cite{LHCb:2021auc,LHCb:2021vvq}. It lies around $360$~keV below the $D^{*+} D^0$ threshold and has a decay width of about $50$~keV. Its favored quantum numbers assignment is $I(J^P) = 0(1^+)$.
The lowest $J^{P}$ multiplets are reported in Table~\eqref{tab:QQqq}. The $T_{cc}(3875)^+$ is identified with the unique member of the lowest $k^P = 0^+$ (or \textit{good diquark}) multiplet.

\begin{table}[h!]
\centering
\resizebox{0.70\columnwidth}{!}{%
	\begin{tabular}{||c|c|c|c|c||c|c||}
		\hline\hline
		\multirow{2}{*}{\hspace{2pt}$\begin{array}{c} QQ\\\text{ color state}\end{array}$\hspace{2pt}} & \multirow{2}{*}{\hspace{2pt}$\begin{array}{c} \text{$\bar{q}\bar{q}$ spin}\\ k^{P}\end{array}$\hspace{2pt}} & \multirow{2}{*}{\hspace{2pt} $\begin{array}{c} \text{BO quantum \#}\\\Lambda^\sigma_\eta\end{array}$\hspace{2pt}} & \multirow{2}{*}{\hspace{2pt}$\begin{array}{c} \text{Isospin}\\I\end{array}$ \hspace{2pt}}& \multirow{2}{*}{\hspace{2pt} $l$\hspace{2pt}} & \multicolumn{2}{|c||}{$J^{P}$}\\
		\cline{6-7}
		& & & & & \hspace{2pt}$S=0$\hspace{2pt} & \hspace{2pt}$S=1$\hspace{2pt} \\
		\hline\hline
		\multirow{4}{*}{\hspace{2pt}$\begin{array}{c} \text{Antitriplet}\\\bar{\mathbf{3}}\end{array}$\hspace{2pt}} &\multirow{2}{*}{$0^+$} & \multirow{2}{*}{${\Sigma_g^+}$} & \multirow{2}{*}{$0$}& $0$ & \hspace{2pt} --- & \hspace{2pt}$1^+$\hspace{2pt} \\
		\cline{5-7}
		& & & & 1 &\hspace{2pt}$1^-$\hspace{2pt}&---\\
		\cline{2-7}
		& \multirow{2}{*}{$1^+$} & \multirow{2}{*}{${\Sigma_g^-, \Pi_g}$} & \multirow{2}{*}{$1$}& $0$ &  \hspace{2pt}$0^-$\hspace{2pt}& --- \\
		\cline{5-7}
		& & & & 1 & $1^{-}$&\hspace{2pt}$\left(0, 1, 2\right)^+$\hspace{2pt}\\
		\cline{5-7}
		\hline\hline
	\end{tabular}}
	\caption{$J^{P}$ multiplets for the lowest $QQ\bar{q}\bar{q}$ tetraquarks with $l\leq1$.
    The dashed entries indicate that that particular state is not allowed due to the Pauli exclusion principle. The multiplets are ordered by increasing energies of the relative states.}
	\label{tab:QQqq}
\end{table}

The BOEFT Schrödinger equation is~\cite{Brambilla:2024imu}
\begin{align}
	\left[-\frac{1}{m_Qr^2}\,\partial_r\,r^2\,\partial_r+\frac{l(l+1)}{m_Qr^2}+V_{\Sigma_{g}^+}(r)\right]\psi_{\Sigma_g^+}^{(N)}=\mathcal{E}_N\,\psi_{\Sigma_g^+}^{(N)}\,.
	\label{eq:SchQQ}
\end{align}
with the $\Sigma_g^+$ static potential, which approaches the $[(QQ)_{\bar{3}} + (\bar{q} \bar{q})_3]$ configuration for $r \rightarrow 0$ and the $[(Q\bar{q})_1 + (Q \bar{q})_1]$ configuration for $r \rightarrow \infty$.
The triplet meson mass $k^P = 0^+$ is tuned in order to generate a single $T_{cc}(3875)^+$ bound state with a large scattering length $a \sim 8$~fm, and around $320$~keV below the $D D^+$ threshold.
If the same equation is solved for the bottom and bc sector, with the $k^P = 0^+$ triplet fixed on the $T_{cc}(3875)^+$ state, we predict two bound states, $T_{bb}$ and $T_{bc}$, respectively $115$~MeV and $25$~MeV below the relative thresholds. These results align well with those of several ab initio lattice studies and of other works adopting the Born-Oppenheimer approximation~\cite{Brambilla:2024imu}.


\section{Threshold effects on the quarkonium spectrum}
\label{sec:quarkonium}

The open-flavor threshold effects on the quarkonium spectrum emerge in BOEFT from the mixing between the quarkonium and the tetraquark potentials sharing the same BO quantum numbers. 
Such threshold effects have been a subject of study since the introduction of prominent phenomenological models, such as the Cornell coupled-channel or the $^3P_0$ model~\cite{Brambilla:2026jqz}. 
The \textit{threshold corrections} on a quarkonium state are defined as
\begin{align}\label{eq:str_break_corr}
\Delta E^{\rm str.br.}_{(N)} = \mathcal{E}_{(N)} - E_{(N)},
\end{align}
where $\mathcal{E}_{(N)}$ is the mass of the state obtained from Eq.~\eqref{eq:coupledSchr}, while $E_{(N)}$ is the corresponding value obtained from the quarkonium channel in the $V_{\Sigma_{g}^{+}-\Sigma_{g}^{+\prime}} \rightarrow 0$ limit.
Alternatively, these can be computed as self-energy corrections to the free quarkonium propagator~\cite{Brambilla:2026jqz}.
These two methods give consistent results for the threshold corrections, whose values are reported for the bottomonium sector in Table~\eqref{tab:coupled_spectra_bb}.
The mixing with tetraquark states induces a systematic downward mass shift for all quarkonium states. The threshold corrections range from about $1$~MeV to $15$~MeV, being negligible for states far from the open-flavor thresholds while increasingly significant for those closer to them. The $4S$ bottomonium state -- corresponding to the spin-averaged $\Upsilon(4S)$ -- acquires a sizable tetraquark admixture of about $20\%$, which may help explain the observed production and decay patterns of this and other similar states.

\begin{table}[ht]
\centering
\renewcommand{\arraystretch}{0.9}
\begin{tabular}{|c||c|c|c|c|c|}  \hline
$n l \, (b \bar{b})$ & $M^{\rm th.}(M^{\rm exp.}) \; (\mathrm{MeV})$ & $\%  \; \Sigma_g^+$ & $\% \; \Sigma_g^{+'}$  & $\% \; \Pi_g$ & $\Delta E^{\rm str. br.}_{nl}\,(\mathrm{MeV})$ \\
\hline\hline
$1S$         & $9444.9 $ $(9445.0)$  & $100$  & $\approx 0$ &   & $-0.1$ \\
$2S$         & $9987.8$ $(10017.3)$  & $99.9$  & $0.1$  &   & $-0.9$\\
$3S$         & $10327.8$ & $99.1$  & $0.9$   &   & $-3.5$\\
$4S$         & $10593.0$  & $77.8$  & $22.2$ &   & $-17.1$\\ \hline
$1P$         & $9882.9 $ $(9899.7)$  & $99.9$  & $0.1$  & $\approx 0$  & $-0.5$\\
$2P$         & $10235.1 $ $(10260.2)$  & $99.6$  & $0.4$  & $\approx 0$ & $-2.2$\\
$3P$         & $10516.1$  & $95.7$  & $4.2$  & $0.1$ & $-7.7$\\
$X_b$        & $10625.9$  & $1.5$  & $44.3$  & $54.2$ & \\\hline
$1D$         & $10124.5$  & $99.8$  & $0.2$  & $\approx 0$  & $-1.1$\\
$2D$         & $10419.0$  & $98.7$  & $1.3$  & $\approx 0$  & $-4.2$\\ \hline
\end{tabular}
\caption{Bottomonium bound state solutions of the spin-averaged Eq.~\eqref{eq:coupledSchr}, denoted by their $nl$ quantum numbers. From the second to the fifth column, mass and composition of the states are reported. The last column shows the threshold corrections.}
\label{tab:coupled_spectra_bb}
\end{table}

The spread among threshold-correction values computed in the literature from phenomenological models, such as the $^3P_0$, stems mainly from their strong sensitivity to the mixing potential, fitted to the spectrum~\cite{Brambilla:2026jqz}.
In BOEFT, instead, all the potentials are unambiguously expressed as generalized Wilson loops and constrained by the available lattice data~\cite{Brambilla:2026jqz}. 

\section{Inclusive hadroproduction of $\chi_{c1}(3872)$ and $X_b$}
\label{sec:hadroprod}

The BOEFT factorization formula allows compute the $\chi_{c1}(3872)$ production cross-section without the need for a fit to prompt hadroproduction data.
The inclusive $\chi_{c1}(3872)$ production cross-section, at leading order in $v$, can be expressed from the NRQCD factorization formula ~\cite{Brambilla:2026ujo}
\begin{align}\label{eq:NRQCDfact}
\sigma_{\chi_{c1}(3872)} = \sigma_{Q\bar{Q}({}^3S_1^{[8]})} \, \langle \Omega | \mathcal{O}^{\chi_{c1}(3872)}({}^3S_1^{[8]}) | \Omega \rangle \, ,
\end{align}
as the product of a known short-distance coefficient (SDC) $\sigma_{Q\bar{Q}({}^3S_1^{[8]})}$ and an unknown long-distance matrix element (LDME) $\langle \Omega | \mathcal{O}^{\chi_{c1}(3872)}({}^3S_1^{[8]}) | \Omega \rangle \,$. 
The LDME can be further factorized:
\begin{align}\label{eq:pNRQCDfact}
 \langle \Omega | \mathcal{O}^{\chi_{c1}(3872)}({}^3S_1^{[8]}) | \Omega \rangle \, = \frac{3}{2 \pi} |\phi_S(0)|^2 \, \mathcal{M}_S \lesssim \frac{2}{\pi} |\phi_S(0)|^2,
\end{align}
as the product of the S-wave $\chi_{c1}(3872)$ wavefunction squared at the origin and a universal matrix element $\mathcal{M}_S$. 
$|\phi_S(0)|^2$ is determined from Eq.~\eqref{eq:coupledSchr}, while  $\mathcal{M}_S$ is extracted from the experimental $\chi_{c1}(3872)$ formation branching ratio in $B$ decays, factorizable as
\begin{align}\label{eq:branching}
\mathrm{Br}(B \to \chi_{c1}(3872) + X) = \mathrm{Br}(b \to c\bar{c}({}^3S_1^{[8]}) + X) \, \langle \Omega | \mathcal{O}^{\chi_{c1}(3872)}({}^3S_1^{[8]}) | \Omega \rangle \, .
\end{align}





\begin{figure}[htb]
    \centering
    \scalebox{0.85}{%
    \begin{minipage}{\textwidth}
        \centering

        \begin{subfigure}[b]{0.43\textwidth}
            \centering
            \includegraphics[width=\textwidth]{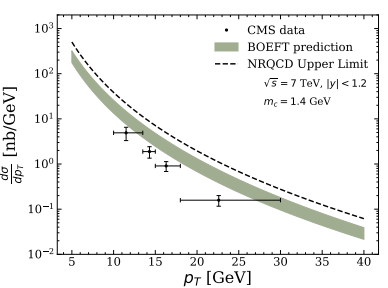}
            \caption{}
            \label{fig:CMS}
        \end{subfigure}
        \hfill
        \begin{subfigure}[b]{0.46\textwidth}
            \centering
            \includegraphics[width=\textwidth]{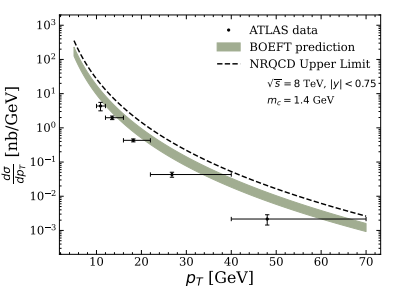}
            \caption{}
            \label{fig:ATLAS}
        \end{subfigure}

        \vspace{0.5cm}

        \begin{subfigure}[b]{0.44\textwidth}
            \centering
            \includegraphics[width=\textwidth]{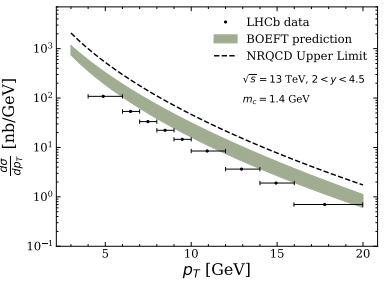}
            \caption{}
            \label{fig:LHCb}
        \end{subfigure}

    \end{minipage}%
    }
    \caption{Theoretical predictions for the prompt inclusive hadroproduction differential cross section of the $\chi_{c1}(3872)$ state as a function of $p_T$~\cite{Brambilla:2026ujo} compared with CMS data~\cite{CMS:2013fpt} (upper-left panel), ATLAS data~\cite{ATLAS:2016kwu} (upper-right panel), and LHCb data~\cite{LHCb:2021ten} (lower panel).}
    \label{fig:chic1_hadroproduction}
\end{figure}

The predictions for the inclusive hadroproduction differential
cross-section compared with the available experimental data are given in Figure~\ref{fig:chic1_hadroproduction}, finding agreement within the given uncertainties.
The same calculation has been extended to the $X_b$ hadroproduction, where only the value of $|\phi_S(0)|^2$ changes, and to the pentaquark production cross-section in~\cite{Brambilla:2026ujo}.


\section{Conclusions}
\label{sec:conclusions}

The Born-Oppenheimer EFT (BOEFT) is a nonrelativistic EFT directly derived from QCD, rooted in the symmetries and the scale separation of the system. It applies to all the XYZ exotic states containing two heavy quarks and any light degrees of freedom, such as hybrids, tetraquarks, and pentaquarks.
The BOEFT potentials appear as matching coefficients and are defined in terms of generalized Wilson loops.
Several lattice calculations have been performed for hybrid potentials, while the tetraquark and pentaquark cases remain largely unexplored.
A precise determination of these potentials is instrumental in discriminating among different proposed scenarios for these exotic states.
Notably, BOEFT makes no prior assumption about their internal structure, e.g. whether a tetraquark is predominantly compact or molecular. Instead, their properties emerge from the solutions of the Schrödinger equations derived within the EFT.

Remerkably, even with only partial knowledge of the static tetraquark potentials, the strong symmetry constraints inherent in BOEFT provide a simple explanation for both the limited number of observed tetraquark candidates and their proximity to meson--antimeson thresholds.
 More precisely, the masses of the associated adjoint mesons determine whether the static potentials approach their asymptotic threshold values from below, thereby allowing for bound-state or resonant solutions, or from above, in which case such solutions do not arise. A precise determination of these quantities from lattice QCD is therefore of pivotal importance for elucidating the underlying pattern of the XYZ exotic spectrum.

In this proceeding, we have discussed several recent applications of BOEFT to prominent exotic candidates.
$\chi_{c1}(3872)$ and $T_{cc}(3875)^+$ emerge as shallow bound states, arising from a conspiracy between the short- and long-range behavior of the potentials.
The open-flavor threshold effects on the quarkonium spectrum are rigorously defined, with the mixing potential expression constrained by existing lattice data, rather than fixed phenomenologically to the spectrum as in the $^3P_0$ model.
Finally, the $\chi_{c1}(3872)$ hadroproduction cross section is computed using the BOEFT factorization formula, finding general agreement, within uncertainties, with the available experimental data.

\section*{Acknowledgements}

I am grateful to Nora Brambilla, Abhishek Mohapatra, and Antonio Vairo for their collaboration on the research presented here.
This work has been supported by the DFG cluster of excellence ORIGINS, funded by the Deutsche Forschungsgemeinschaft under Germany’s Excellence Strategy-EXC-2094-390783311.
I also acknowledge the ERC Advanced Grant ERC-2023-ADG-Project EFT-XYZ.

\bibliography{bibliography}

\end{document}